\documentclass[conference, letterpaper]{IEEEtran}
\ifCLASSINFOpdf
\else
\fi
\usepackage{url}
\usepackage{subcaption}

\ifCLASSINFOpdf
   \usepackage[pdftex]{graphicx}
\else
\fi

\DeclareGraphicsExtensions{.pdf,.jpeg,.png}
\usepackage[cmex10]{amsmath}
\usepackage{color}
\usepackage{fancyhdr}

\renewcommand{\thispagestyle}[2]{}

\fancypagestyle{plain}{
        \fancyhead{}
        \fancyhead[C]{first page center header}
        \fancyfoot{}
        \fancyfoot[C]{first page center footer}
}
\begin{document}

\title{Open Algorithms for Identity Federation}

\author{\IEEEauthorblockN{Thomas Hardjono}
\IEEEauthorblockA{Connection Science \& Media Lab\\
Massachusetts Institute of Technology\\
Cambridge, MA 02139, USA\\
Email: hardjono@mit.edu}
\and
\IEEEauthorblockN{Alex Pentland}
\IEEEauthorblockA{Connection Science \& Media Lab\\
Massachusetts Institute of Technology\\
Cambridge, MA 02139, USA\\
Email: sandy@media.mit.edu}}


%Telephone: (800) 555--1212, Fax: (888) 555--1212}


\maketitle

\begin{abstract}
%\boldmath
The identity problem today is a data-sharing problem.
Today the fixed attributes approach adopted
by the consumer identity management industry
provides only limited information about an individual,
and therefore is of limited value to the service providers
and other participants in the identity ecosystem.
This paper proposes the use of the {\em Open Algorithms} (OPAL) paradigm
to address the increasing need for individuals and 
organizations to share data in a privacy-preserving manner.
Instead of exchanging static or fixed attributes,
participants in the ecosystem
will be able to obtain better insight through a collective {\em sharing of algorithms},
governed through a trust network.
Algorithms for specific data-sets must be vetted to be privacy-preserving,
fair and free from bias.

\end{abstract}



% \fi
\IEEEpeerreviewmaketitle

%%%% ---------- Section --------------------
\section{Introduction}
\label{sec:intro}

The Open Algorithms (OPAL) paradigm seeks
to address the increasing need for individuals and 
organizations to share data in a privacy-preserving manner~\cite{PentlandShrier2016}.
Data is crucial to the proper functioning of communities,
businesses and government.
Previous research has indicated that data
increases in value when it is combined.
Better insight is obtained when different types of data
from different areas or domains
are combined.
These insights allows communities to begin addressing 
the difficult social challenges of today,
including better urban design,
containing the spread of diseases,
detecting factors that impact the economy,
and other challenges of the data-driven society~\cite{Pentland2015,WISH2013}.

Today there are a number of open challenges with regards to
the data sharing ecosystem:
\begin{itemize}

\item	{\em Data is siloed}: Today data is siloed within organizational boundaries,
and the sharing of raw data with parties outside the organization remains unattainable,
either due to regulatory constraints or due to business risk exposures. 

\item	{\em Privacy is inadequately addressed}: The 2011 World Economic Forum (WEF) report~\cite{WEF2011} 
on personal data as a new asset class
finds that the current ecosystems that access and use personal data is fragmented and inefficient.
For many participants, the risks and liabilities exceed the
economic returns and personal privacy concerns are inadequately addressed.
Current technologies and laws fall short of providing
the legal and technical infrastructure needed to
support a well-functioning digital economy.
The rapid rate of technological change and commercialization
in using personal data is undermining end-user
confidence and trust.

\item	{\em Regulatory and compliance}:
The introduction of the EU General Data Protection Regulations (GDPR)~\cite{GDPR}
will impact global organizations that rely on the Internet 
for trans-border flow of raw data.
This includes cloud-based processing sites that are spread across the globe.

\end{itemize}

Similarly, today there are a number of challenges
in the identity and access management space,
notably in the area of consumer identity:
\begin{itemize}

\item	{\em Identity tied to specific services}:
Most digital ``identities'' (namely identifier strings
such as  email addresses) are created as an 
adjunct construction to support access to specific services on the Internet. 
This tight coupling between digital identifiers and services 
has given rise to the unmanageable proliferation 
of user-accounts on the Internet. 

\item	{\em Massive duplication of data}:
Together with the proliferation of user-accounts comes 
the massive duplication of personal attributes across 
numerous service providers on the Internet. 
These service providers are needlessly holding 
the same set of person-attributes
(e.g. name, address, phone, etc.)
associated with a user.

\item	{\em Lack or absence of user control}:
In many cases users have little knowledge about what data
is collected by service provider,
how the data was collected and the actions taken on the data.
As such, end-users have no control over the other usages
of their data beyond what was initially consented to.

\item	{\em Diminishing trust in data holders or custodians}:
The laxity in safeguarding user data has
diminished social trust on the part of users in
entities which hold their data.
Recent attacks on data repositories and 
theft of massive amounts of data
(e.g. Anthem hack~\cite{Anthem2015}, Equifax attack~\cite{Equifax2017}, etc.)
illustrates this ongoing problem.

\item	{\em Misalignment of incentives}:
Today customer-facing service providers (e.g. online retail)
have access only to poor quality user data.
Typically such data is
obtained from data aggregators
who in turn collate an incomplete picture of the user
through various back-channel means (e.g. ``scraping'' various Internet sites).
The result is a high cost to service providers for new customer on-boarding,
coupled with low predictive capabilities of the data.

\end{itemize}

The identity problem today is in reality a data-sharing problem.
The overall goal of this paper is to provide an alternate
architecture for identity management
based on the open algorithms paradigm.
Key to this approach is the notion of sharing information
in a privacy-preserving manner
based on vetted algorithms, instead of the exchange
of fixed attributes approach that has prevailed
in the identity industry for the past two decades.

The remainder of the paper is arranged as follows.
Section~\ref{sec:id-federation} provides a brief history and overview
of identity management and federation,
providing some definitions of the entities and their functions.
Readers familiar with the IAM industry and the current
identity federation landscape can skip this section.

Section~\ref{sec:concepts} provides further detail the concepts and principles
underlying the open algorithms paradigm.
Section~\ref{sec:opalidentityfederation} addresses the open algorithms
paradigm in the context of identity federation,
while
Section~\ref{sec:opal-trust-framework} briefly discusses the need
for a legal trust framework to share algorithms.
Section~\ref{sec:opal-Consent} briefly touches on the topic of subject consent.

The paper closes with a discussion regarding possible future directions
for the open algorithms paradigm.

%%%% ---------- Section --------------------

\section{Identity Federation and Attribute Sharing:\\ A Brief History}
\label{sec:id-federation}

\begin{figure}[!t]
\centering
	% ORIGINAL \includegraphics[width=3.0in]{hardjono-classic-id-federation-v02-pdf}
	%
\includegraphics[width=0.4\textwidth, trim={0.5cm 0.5cm 0.5cm 0.5cm}, clip]{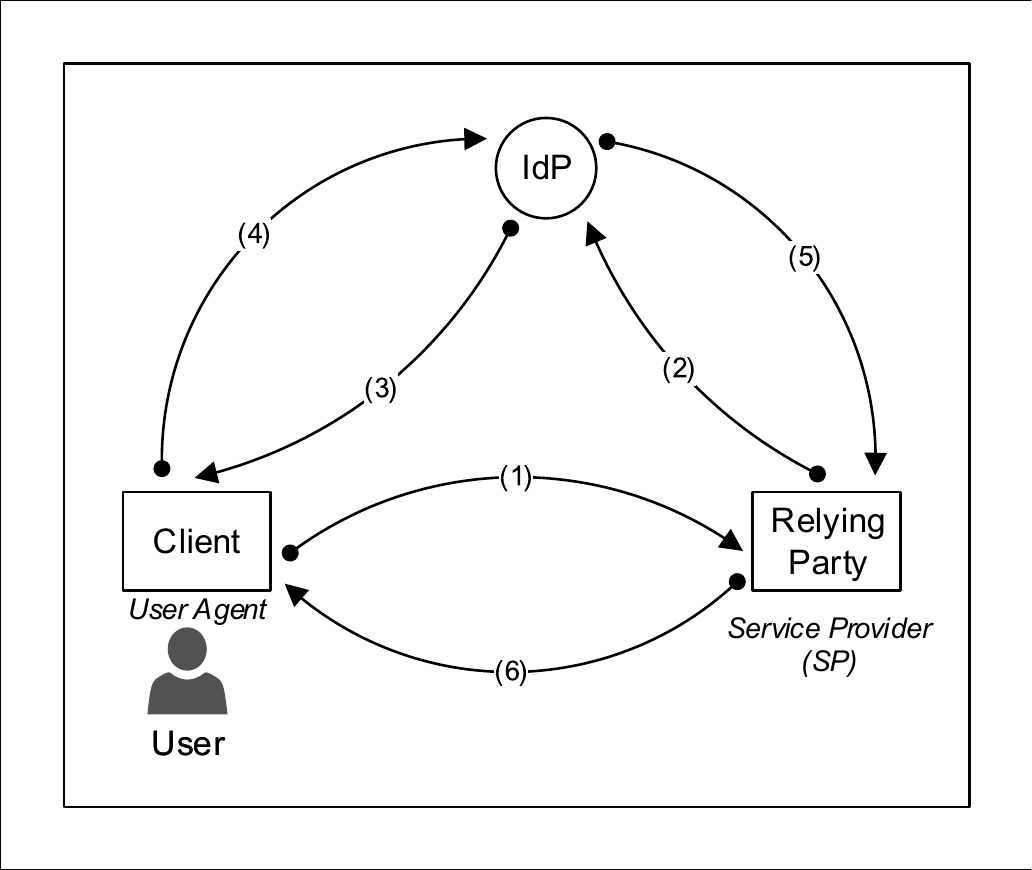}
	%
	% where an .eps filename suffix will be assumed under latex, 
	% and a .pdf suffix will be assumed for pdflatex; or what has been declared
	% via \DeclareGraphicsExtensions.
	%
	% TRIMMING:  trim={<left> <lower> <right> <upper>} and clip options:
	% FULL EXAMPLE: \includegraphics[width=0.4\textwidth, trim={0.5cm 0.5cm 0.5cm 11.3cm}, clip]{image1.pdf}
	%
\caption{Overview of Web Single Sign On (Web-SSO)}
\label{fig:web-sso}
\end{figure}

Today {\em Identity and Access Management} (IAM) 
represents a core component of the Internet.
IAM infrastructures are an enabler 
which allows organizations to achieve its goals.
Enterprise-IAM (E-IAM) is already a mature product category~\cite{Gartner2016a} and
E-IAM systems are already well integrated into
other enterprise infrastructures, such as directory services
for managing employees and corporate assets.
The primary goal of E-IAM systems is to authenticate and identify
persons (e.g. employees) and devices as they enter the enterprise boundary,
and to enforce access control driven by corporate policies.

In the case of Consumer-IAM (C-IAM) systems, the primary goal there
is to reduce friction between the consumer and the online service provider (e.g. merchant)
through a mediated-authentication process,
using a trusted third party referred to as the {\em Identity Provider}.

\subsection{Web Single Sign-On \& Identity Providers}

Historically, notion of the {\em identity provider} entity emerged starting in the late 1990s
in response to the growing need to aid users in accessing services online.
During this early period,
a user would typically create an account and credentials at every new service provider (e.g. online merchant).
This cumbersome approach, which is still in practice today,
has led to a proliferation of accounts 
and duplication of the same user attributes across many service providers.

The solution that emerged became what is known today as {\em Web Single Sign-On} (Web-SSO)~\cite{SAMLwebsso}.
The idea is that a trusted third party referred to as the {\em Identity Provider} (IdP)
would mediate the authentication of the user on behalf of the service provider.
This is summarized in Figure~\ref{fig:web-sso}.
When a user visits a merchant website (Step~1), the user would be redirected to the IdP 
to perform authentication (Steps~2--4).
After the completion of a successful authentication event,
the IdP would redirect the user's browser back to the calling merchant (Step~5),
accompanied by an IdP-signed assertion stating that the IdP has successfully
authenticated the user.
The merchant would then proceed transacting with the user (Step~6).

In order to standardize this protocol behavior,
in 2001 an alliance of over 150 companies and organizations formed an industry consortium
called the {\em Liberty Alliance Project}.
The main goal of this consortium was to ``establish standards, guidelines and best 
practices for identity management''~\cite{ProjectLiberty}.
Several significant outcomes for the IAM industry
resulted from Liberty Alliance, two among which were: 
(a) the standardization of 
the {\em Security Assertions Markup Language} (SAML2.0)~\cite{SAMLcore},
and (b) the creation of a widely used open source {SAML2.0} 
server implementation called {\em Shibboleth}~\cite{Shibboleth2004}.

Today {SAML2.0} remains the predominant Web-SSO technology used within Enterprise IAM,
which is directly related to the type of authentication protocol dominant
in enterprise directory services~\cite{rfc4120,rfc4178,rfc4559}.

In the Consumer IAM space, 
the growth of the web-applications industry has spurred the creation
of the {OAuth2.0} framework~\cite{rfc6749} based on JSON web tokens.
Similar to the SSO pattern,
the purpose of this framework is to authenticate and authorize
an ``application'' to access a user's ``resources''.
The {OAuth2.0} model follows the traditional notion of {\em delegation}
where the user as the resource-owner authorizes an application, 
such as a Web Application or Mobile Application,
to access the user's resources (e.g. files, calendar, other accounts, etc).
In contrast to {SAML2.0} which requires the user to be present at the browser
to interact with the service provider,
in {OAuth2.0} the user can disconnect after he or she authorizes
the application to access his or her resources.
In effect the user is delegating control to the application (to perform some defined task)
in the absence of the user.

Although the {OAuth2.0} design as defined in~\cite{rfc6749} lacks details
for practical implementation and deployment, 
a fully fledged system is defined in the {OpenID-Connect 1.0} protocol~\cite{OIDC1.0} specification
based on the {OAuth2.0} model.
It is this {OpenID-Connect 1.0} protocol (or variations of it) which is today deployed
by the major social media platforms in the Consumer-IAM space.

\begin{figure}[!t]
\centering
	% ORIGINAL \includegraphics[width=3.0in]{hardjono-classic-id-federation-v02-pdf}
	%
\includegraphics[width=0.4\textwidth, trim={0.5cm 0.5cm 0.5cm 0.5cm}, clip]{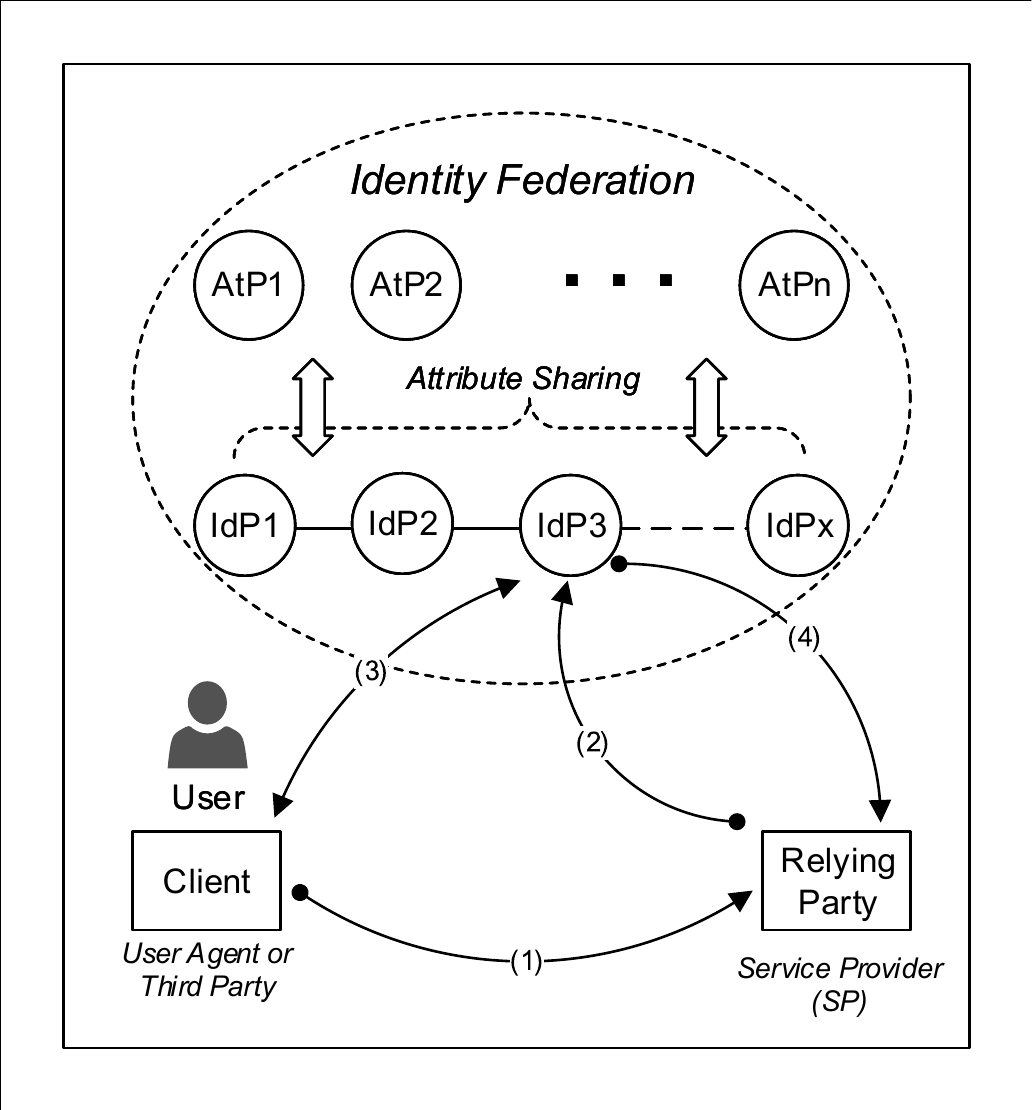}
	%
	% where an .eps filename suffix will be assumed under latex, 
	% and a .pdf suffix will be assumed for pdflatex; or what has been declared
	% via \DeclareGraphicsExtensions.
	%
	% TRIMMING:  trim={<left> <lower> <right> <upper>} and clip options:
	% FULL EXAMPLE: \includegraphics[width=0.4\textwidth, trim={0.5cm 0.5cm 0.5cm 11.3cm}, clip]{image1.pdf}
	%
\caption{Identity Federation and Attribute Sharing}
\label{fig:traditional-sso}
\end{figure}

\subsection{Identity Federation}

The notion of {\em federation} among identity providers arose
from a number of practical needs,
one being that of the scaling of services.
The idea is quite straightforward and extends from the previous scenario
of a user seeking access to a service provider or relying party (i.e. online merchant).
This is shown in Figure~\ref{fig:traditional-sso}.
The problem was simply the following:
when the relying party directs the user to an IdP with whom the relying party has a business relationship,
there was a chance that the user will be unknown to that IdP.
As such, the solution is for a group of IdPs to ``network together''
in such a way that when one IdP is faced with an unknown user, 
the IdP can inquire with other IdPs in the federation.
The federation model opens-up other interesting possibilities,
including the possible introduction of the so-called {\em attribute provider} (AtP) entity
whose primary task is to issue additional useful assertions about the user.

\begin{figure*}[!t]
\centering
	% ORIGINAL \includegraphics[width=3.0in]{hardjono-classic-id-federation-v02-pdf}
	%
\includegraphics[width=0.7\textwidth, trim={0.5cm 0.5cm 0.5cm 0.5cm}, clip]{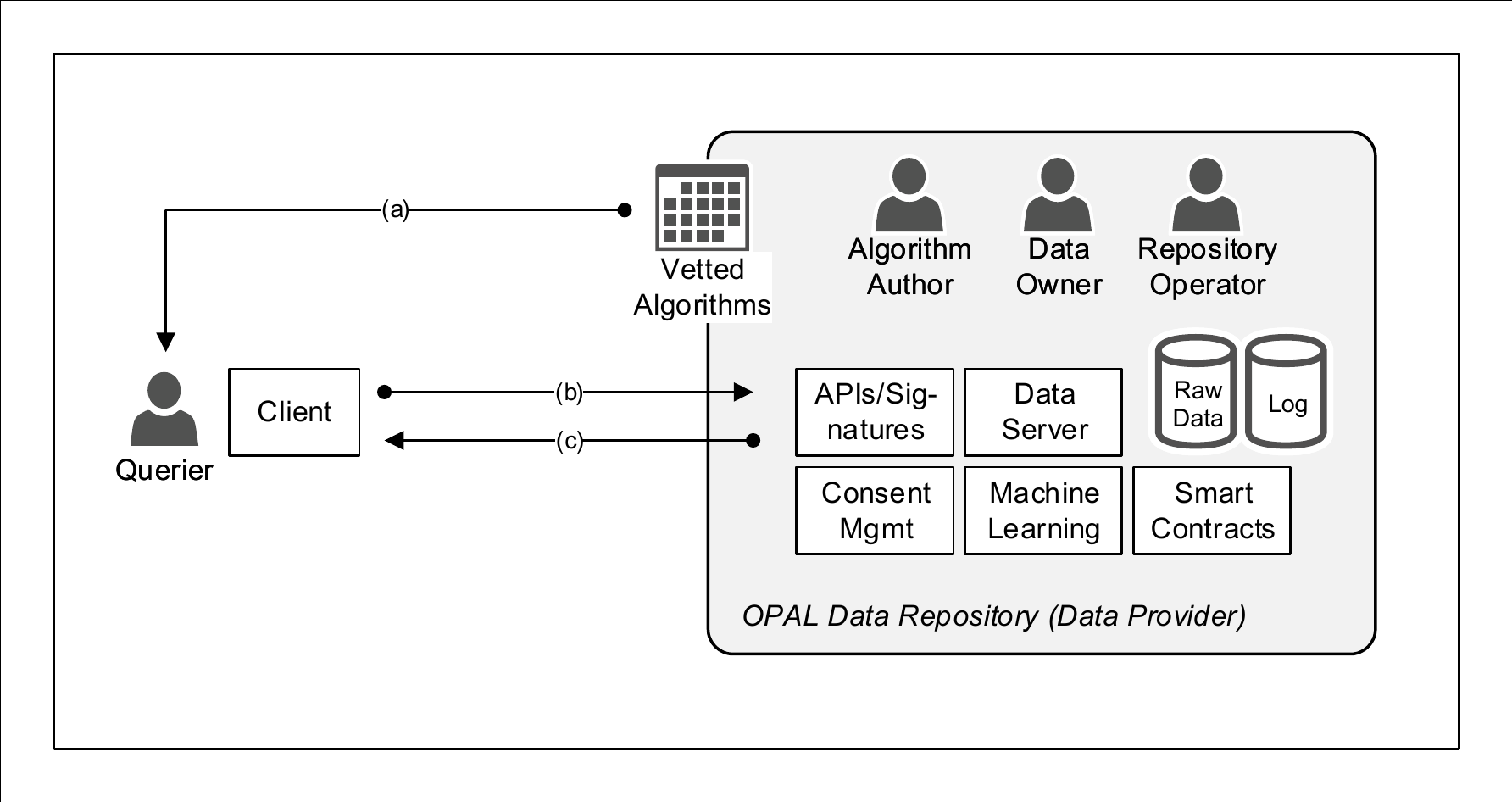}
	%
	% where an .eps filename suffix will be assumed under latex, 
	% and a .pdf suffix will be assumed for pdflatex; or what has been declared
	% via \DeclareGraphicsExtensions.
	%
	% TRIMMING:  trim={<left> <lower> <right> <upper>} and clip options:
	% FULL EXAMPLE: \includegraphics[width=0.4\textwidth, trim={0.5cm 0.5cm 0.5cm 11.3cm}, clip]{image1.pdf}
	%
\caption{Overview of Open Algorithms (OPAL) Architecture}
\label{fig:opalarchitecture}
\end{figure*}

More formally, the primary goal of a {\em federation} among a group of {\em identity providers} (IdP)
is to share ``attribute'' information (assertions) regarding a user~\cite{ABA2012,SAMLglossary}:
\begin{itemize}

\item	An {\em Identity Provider} is the entity (within a given identity system) which is 
responsible for the
identification of persons, legal entities, devices, and/or digital objects, the issuance of
corresponding identity credentials, and the maintenance and management of such identity
information for Subjects.

\item	An {\em attribute} is a specific category of 
identifying information about a subject or user.
Examples for users include name,
address, age, gender, title, salary, health, net worth, driver's license number, Social
Security number, etc.

\item	An {\em Attribute Provider} (AtP) is a  third party trusted as an authoritative source of information and
responsible for the processes associated with establishing and maintaining identity
attributes. An Attribute Provider asserts trusted, validated attribute claims in response to
attribute requests from Identity Providers and Relying Parties. Examples of Attribute
Providers include a government title registry, a national credit bureau, or a commercial
marketing database.

\item	An {\em Identity Federation} is the set of technology, standards, policies, and processes that allow an
organization to trust digital identities, identity attributes, and credentials created and issued
by another organization. A federated identity system allows the sharing of identity
credentials issued, and identity information asserted, by one or more Identity Providers
with multiple Relying Parties (RP).

\item	A  {\em Relying Party or Service Provider} (RP) is  system entity that decides to take an action based on
information from another system entity. For example, a SAML
relying party depends on receiving assertions from an authoritative asserting
party (an identity provider) about a subject~\cite{SAMLglossary}.

\end{itemize}

Although the federated identity model using the SSO flow pattern 
remains the predominant model today for the consumer space,
there are a number of limitations of the model -- both
from the consumer privacy perspective and from 
the service providers business model perspective:

\begin{itemize}

\item	{\em Identity management as an adjunct service}: Most large scale consumer-facing
identity services today are a side function
to another more dominant service (e.g. free email service, free social media, etc.).

\item	{\em Limited access to quality data}: Service providers and relying parties
are seeking better insights into the user,
beyond the basic attributes of the user.
However, other than the major social media providers today,
the relying parties today do not have access to rich data regarding the user.

~~Today the scope of attributes or claims being exchanged among federated identity providers
is fairly limited (e.g. user's name, email, telephone, etc.)
and thus of little value to the relying party.
An example of the list of claims can be found
in the OpenID-Connect {1.0} core specifications (Section {5.1} on Standard Claims in~\cite{OIDC1.0})
for federations deploying the OpenID-Connect architecture.

\item	{\em Limited user control and consent}: Today the user is typically
``out of the loop'' with regards to consent regarding the information being asserted to
by an identity provider or attribute provider.

~~Over the past few years
several efforts have sought to address the issue of user control and consent (e.g.~\cite{UMACORE1.0,UMACORE2.0,CONSENT1.0}).
The idea here is that not only should the user explicitly consent to his or her attributes being shared,
the underlying identity system should also ensure that only minimal disclosure is performed for a constrained use
among justifiable parties~\cite{Cameron2004identity,Cavoukian2006}.

\end{itemize}

%%%% ---------- Section --------------------
\section{Open Algorithms: Key Concepts \& Principles}
\label{sec:concepts}

The concept of {\em Open Algorithms} (OPAL) evolved
from several research projects over the past decade
within the Human Dynamics Group at the MIT Media Lab.
From various research results
it was increasingly becoming apparent that an individual's privacy
could be affected through the correlation of just small 
amounts of data~\cite{Montjoye2013,pentland-saving-big-data-2014}.

One noteworthy seed project was {\em OpenPDS} that sought
to develop further the concept of personal data stores (PDS)~\cite{hardjono1996a,openPDS2014PLOS},
by incorporating the idea of analytics on personal data and the notion
of ``safe answers'' as being the norm for responses generated by a personal data store.

However, beyond the world of personal data stores
there remains the pressing challenges around how
large data stores are to secure their data,
safeguard privacy and  comply to regulations (e.g. GDPR~\cite{GDPR}) --
while at the same time enable productive collaborative data sharing.
The larger the data repository, the more attractive it would become to hackers
and attackers.
As such, it became evident that the current mindset of performing data analytics
at a centralized location needed to be replaced with
a new paradigm for thinking about data sharing in a distributed manner.

The OPAL paradigm provides a useful framework
within which industry can begin finding solutions for these constraints.

\subsection{OPAL Principles}

The following are the key concepts and principles underlying the 
open algorithms paradigm~\cite{PentlandShrier2016}:
\begin{itemize}

\item	{\em Moving the algorithm to the data}: Instead of pulling raw data into
a centralized location for processing, it is the algorithms that 
should be sent to the data repositories and be processed there.

\item	{\em Raw data must never leave its repository}: Raw data must never be exported
from its repository, 
and must always be under the control of its owner.

\item	{\em Vetted algorithms}:  Algorithms must be vetted 
to be ``safe'' from bias, discrimination, privacy violations and other unintended consequences.
The data owner (data provider) must ensure that the algorithms which it publishes
have been thoroughly analyzed for safety and privacy-preservation.

\item	{\em Provide only safe answers}: When executing an algorithm on a data-set,
the data repository must always provide responses that
are deemed ``safe'' from a privacy perspective. 
Responses must not release personally identifying information (PII)
without the consent of the user (subject).
This may imply that a data repository return {\em only aggregate answers}.

\item	{\em Trust Networks (Data Federation)}: In a group-based information sharing 
configuration -- referred to as {\em Trust Network for Data Sharing Federation} --
algorithms must be vetted collectively by the trust network members.
Individually, each member must observe the OPAL principles and operate on this basis.
The operational aspects of the federation
should be governed by a {\em legal trust framework} (see Section~\ref{sec:opal-trust-framework}).

\item	{\em Consent for algorithm execution}: Data repositories that hold subject 
data must obtain explicit consent from the subject
when this data is to be included in a given algorithm execution.

~~This implies that as part of obtaining a subject's consent,
the vetted algorithms should be made available and understandable to subjects.
Consent should be unambiguous and retractable (see Article~7 of GDPR~\cite{GDPR}).
Standards for user-centric authorization and consent-management~\cite{UMACORE1.0,CONSENT1.0}
exist today in the identity industry, and which can be the basis
for managing subject consent in data repositories.

\item	{\em Decentralized Data Architectures}: By leaving raw data in its repository,
the OPAL paradigm points towards a decentralized architecture for data stores~\cite{pentland-saving-big-data-2014}.
Decentralized data architectures based
on standardized interfaces/APIs
should be applicable to personal data stores as legitimate end-points.
That is, the OPAL paradigm should be applicable regardless of the size of the data-set.

~~New architectures based on Peer-to-Peer (P2P) networks should be employed
as the basis for new decentralized data stores~\cite{DeFilippiMcCarthy2012}.
Since data is a valuable asset, the proper design of a decentralized architecture
should aim at increasing the resiliency of the overall system against attacks 
(e.g. data theft, poisoning, manipulations, etc).
New distributed data security solutions
based on secure multi-party computation 
(e.g. MIT~Enigma~\cite{ZyskindNathan2015})
and homomorphic encryption, should be further developed.

~~Additionally, a decentralized service architecture
should enhance distributed data stores.
Such a service architecture should introduce automation in
the provisioning and de-provisioning of services
through the use of smart contracts technology~\cite{Hardjono2017a}.

\end{itemize}

It is important to note
that the term ``algorithm'' itself has been left intentionally undefined.
This is because each OPAL deployment instance must have the flexibility
to define the semantics and syntax of their algorithms.
In the case of a community of data providers organized under a trust network,
they must collectively agree on the semantics and syntax
in the operational sense.
Such a definition should be a core part of the legal trust framework
underlying the federated community.

\subsection{OPAL Query-Response Model}
\label{subsec:opal-query-response}

From a technological perspective, 
the OPAL model is fairly simple to understand (see Figure~\ref{fig:opalarchitecture}).
A querier entity (e.g. person or organization) that wishes to obtain information from 
a given data provider selects one or more vetted algorithms (Step~(a)).

In the MIT software implementation of OPAL,
each algorithm is encapsulated in a ``template'' format that contains the
algorithm description,
its algorithm-ID,
the identifier of the intended (target) repository,
the data-schema,
and other parameters (e.g. cost to querier).
The algorithm template itself is digitally signed (e.g. by the data provider or algorithm author)
to ensure the source-authenticity of the template.

For a given algorithm,
Querier uses this template to construct an ``OPAL contract'' (a JSON data structure)
which contains among others the desired Algorithm-ID.
The contract is digitally signed by the Querier and sent to the target data repository.
This is shown as Step~(b) in Figure~\ref{fig:opalarchitecture}.
Optionally the querier may attach payment in order
to remunerate the data repository.

The data repository validates the signature on the OPAL contract,
checks the identity of the Querier
and executes the algorithm (corresponding to the Algorithm-ID)
on the target data-set in its back-end.

The results are then placed into a ``OPAL contract-response'' (another JSON data structure)
by the data provider, digitally signed and returned to the Querier.
This is shown as Step~(c) in Figure~\ref{fig:opalarchitecture}.
Optionally, if confidentiality of the query/response is required then
the relevant entries in the OPAL contract and contract-response could be encrypted.

\subsection{OPAL-based Data Sharing Federation}

\begin{figure*}[!t]
\centering
	% ORIGINAL \includegraphics[width=3.0in]{hardjono-classic-id-federation-v02-pdf}
	%
\includegraphics[width=0.7\textwidth, trim={0.5cm 0.5cm 0.5cm 0.5cm}, clip]{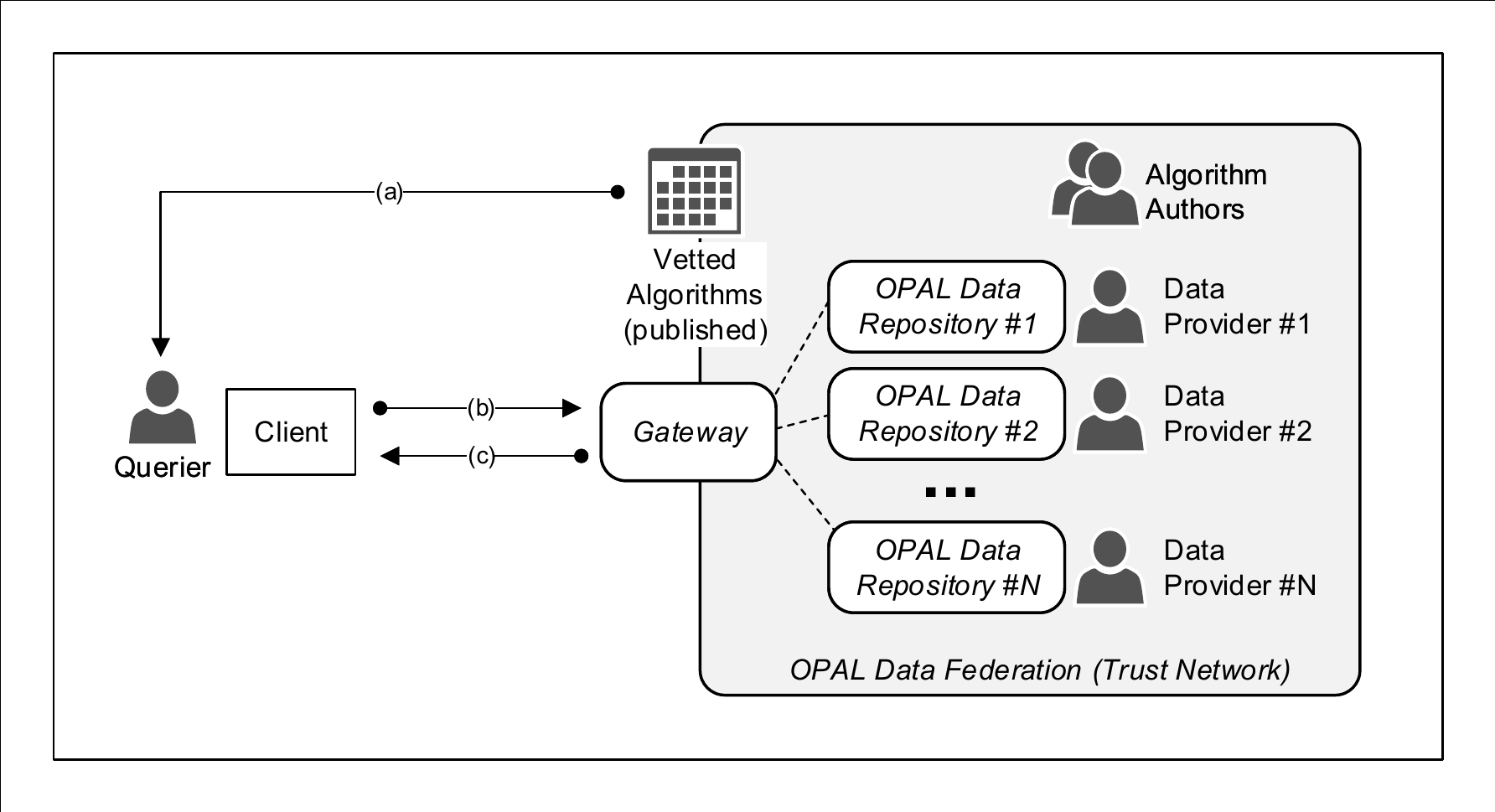}
	%
	% where an .eps filename suffix will be assumed under latex, 
	% and a .pdf suffix will be assumed for pdflatex; or what has been declared
	% via \DeclareGraphicsExtensions.
	%
	% TRIMMING:  trim={<left> <lower> <right> <upper>} and clip options:
	% FULL EXAMPLE: \includegraphics[width=0.4\textwidth, trim={0.5cm 0.5cm 0.5cm 11.3cm}, clip]{image1.pdf}
	%
\caption{Data Sharing Federation using OPAL}
\label{fig:opaldatafederation}
\end{figure*}

In a data sharing federation configuration (e.g. consortium of data providers),
the federation may employ a {\em gateway} entity that coordinates
queries/responses they receive from each other (or from outside
if the federation permits).
This is shown in Figure~\ref{fig:opaldatafederation}.

Here the gateway entity mediates requests coming from the querier entity
and directs them to the appropriate member data provider.
The gateway also collates responses and packages the responses prior to transmitting
to the querier.
Note that a member data provider may always decline to respond
(e.g. data unavailable, it detects attempts to send multiple related queries, etc.).

Currently, small test-bed deployments of the basic open algorithms concept 
are underway for specific and narrow data-domains~\cite{FreyHardjono2017,DataPop-website}.

%%%% ---------- Section --------------------
\section{OPAL for Identity Federation}
\label{sec:opalidentityfederation}

\begin{figure*}[!t]
\centering
	% ORIGINAL \includegraphics[width=3.0in]{hardjono-classic-id-federation-v02-pdf}
	%
\includegraphics[width=0.8\textwidth, trim={0.5cm 0.5cm 0.5cm 0.5cm}, clip]{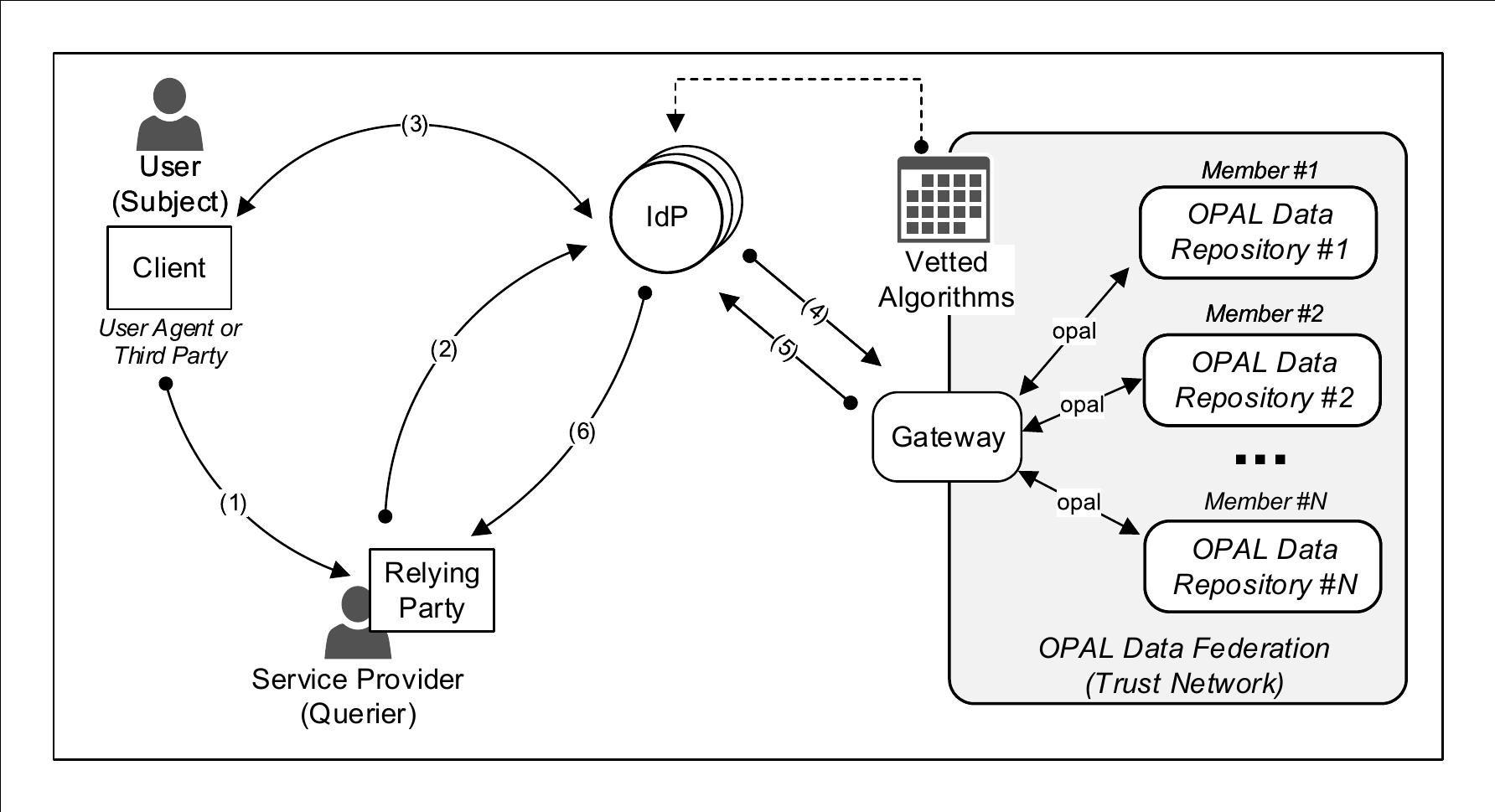}
	%
	% where an .eps filename suffix will be assumed under latex, 
	% and a .pdf suffix will be assumed for pdflatex; or what has been declared
	% via \DeclareGraphicsExtensions.
	%
	% TRIMMING:  trim={<left> <lower> <right> <upper>} and clip options:
	% FULL EXAMPLE: \includegraphics[width=0.4\textwidth, trim={0.5cm 0.5cm 0.5cm 11.3cm}, clip]{image1.pdf}
	%
\caption{OPAL-based Identity and Data Federation}
\label{fig:opalidentityfederation}
\end{figure*}

Instead of the exchange of static attributes regarding a user or subject,
identity providers and data providers should collectively share vetted algorithms
following the open algorithms paradigm.
This is summarized in Figure~\ref{fig:opalidentityfederation}.
\begin{itemize}

\item	{\em Algorithms instead of attributes}: Rather than 
delivering static attributes (e.g. ``Joe is a resident of NY City'') to the relying party,
allow instead the relying party to choose one or more vetted algorithms
from a given data domain (Step~(a)).

~~~The result from executing a chosen algorithm is then conveyed by the IdP
to the relying party in a signed response (Step~(c)).
The response can also include various metadata embellishments,
such as the duration of the validity of the response (e.g. for dynamically changing data sets),
identification of the data-sets used,
consent-receipts, timestamps, and so on.

\item	{\em Convergence of Federations}: Identity federation networks should
engage data provider networks (data owners) in a collaborative manner,
with the goal of converging the two types of networks.

~~The goal should be to bring together data providers from differing
domains or verticals (e.g. telco data, health data, finance data, etc)
in such a manner that new insights can be derived about a subject
with their consent based on the OPAL paradigm.
Research has shown, for example, that combining location data with 
credit card data offers new insights into the financial wellbeing of a user~\cite{MoneyWalks2015}.

\item	{\em Apply correct pricing models for algorithms and data}: 
For each algorithm and the data to which the algorithm applies,
a correct pricing structure needs to be developed by the members of the federation.
This is not only to remunerate the data repositories
for managing the data-sets and for executing the algorithm (i.e CPU usage),
but also to encourage data owners to develop new business models
based on the OPAL paradigm. 

~~~Pricing information could published as part of the vetted-algorithms declaration
(e.g. as metadata), offering different tiers of pricing for different sizes of data sets.
For example, the price for obtaining insights into the creditworthiness of a subject 
based solely on their credit card data
should be different from the price for obtaining insights based on combined data-sets
across domains
(e.g. appropriate combination of GPS data-set and credit card data-set)~\cite{Pentland2015}.

\item	{\em Remunerate the subjects}: 
A correct alignment of incentives must be found 
for all stakeholders in the identity federation ecosystem.
Subjects should see some meaningful and measureable return on the use of their data,
even if it is tiny amounts (e.g. in the pennies or sub-pennies).
Returns should be measurable against some measure of data
the subject puts forward (e.g. variety of data, duration of collection, etc).

~~~The point here is that people will contribute more personal data
if they are active participants in the ecosystem
and understand the legal protections offered 
by the trust frameworks that govern the data federation
and govern the treatment of their personal data by member data providers.

\item	{\em Logs for Transparency and Regulatory Compliance}: All requests and responses
must be logged, together with strong audit capabilities.
Emerging technologies such as blockchains and distributed ledgers
could be expanded to effect an efficient but immutable log of events.
Such a log will be relevant for post-event auditing
and for proving regulatory compliance.

~~~Logging and audit is also crucial in order to obtain social acceptance and consent from individuals
whose data is present within a data repository.
Users as stakeholders in the ecosystem must be able to see
what data is present within these repositories
and to see who is accessing their data through
the execution of vetted algorithms.

\end{itemize}

The use of the OPAL paradigm for information sharing within an identity federation
is summarized in Figure~\ref{fig:opalidentityfederation}
using the traditional Web-SSO flows.
Figure~\ref{fig:opalidentityfederation} shows an alternate flow pattern,
which essentially replaces the attribute providers in 
Figure~\ref{fig:traditional-sso} with the OPAL-based federation.

In Step~1 of Figure~\ref{fig:opalidentityfederation}
when the user seeks the services of the relying party,
the relying party has the option
to request the execution of one or more of the vetted-algorithms
as part of the redirection of the user to the IdP for authentication (Step~2).
Thus, in addition to performing user-authentication the IdP
would forward requests for algorithm execution pertaining to the user (as subject)
to the data providers as federation members (Step~4).
Data providers whose repositories contain data relevant to the relying party
could respond to this request from the IdP (Step~5).
The IdP then relays these signed OPAL responses to the relying party (Step~6).

Note that in Figure~\ref{fig:opalidentityfederation} the relying party could in fact
bypass the IdP by going straight to the Gateway and the Data Federation.
In this topology, the relying party would become the querier and the Gateway itself could in fact
play the dual role of also being an IdP.

%%%% ---------- Section --------------------
\section{Trust Framework for OPAL Federation}
\label{sec:opal-trust-framework}

Today trust frameworks for identity management and federation in the US
is based on three types or ``layers'' of law~\cite{OIX2017}.
The foundational layer is the general commercial law
that consists of legal rules that are applicable to identity management systems and transactions.
This general commercial law was not created specifically for identity management,
but instead are public law written and enacted by governments which applies to all identity systems,
its participants -- and thus enforceable in courts.

The second ``layer'' consist of general identity system law.
Such law is written to govern all identity systems within its scope.
The intent would be to address the various issues related to the operations of the identity systems.
The recognition of the need for law at this layer is new,
perhaps reflecting the slow pace of development in the legal arena as compared to
the technology space.
An example of this is the Virginia Electronic Identity Management Act~\cite{VirginiaLaw2015},
which was enacted recently.

The third ``layer'' is the set of applicable legal rules and system-specific rules
(i.e. specific  to the identity system in question).
The term ``trust framework'' is often used to refer to these
system rules that have been adopted by the community.
A trust framework is needed for a group of entities to govern their collective behavior,
regardless if the identity system is operated by the 
government or the private sector.
In the case of a private sector identity system,
the governing body consisting of the participants in the system
typically drafts rules that take the form of {\em contracts-based rules},
based on private law.

Examples of trust frameworks for identity federation today are FICAM for federal employees~\cite{FICAM},
SAFE-BioPharma Association~\cite{SAFE-BioPharma-2016} for the biopharmaceutical industry,
and the OpenID-Exchange (OIX)~\cite{OIX2017}
for federation based on the {OpenID-Connect 1.0} model.

In order for an OPAL-based information sharing federation to be developed,
it should use and expand the current existing legal trust frameworks for identity systems.
This is because the overall goal is for entities to obtain richer information regarding
a user (subject), and as such it must be bound to the specific identity system rules.
In other words, a new set of third layer legal rules and system-specific rules
must be devised that must clearly articulate the required combination of technical
standards and systems, business processes and procedures, and legal rules that, taken
together, establish a trustworthy system~\cite{ABA2012} for information sharing based on the OPAL model.
It is here that system-specific rules regarding the ``amount of private information released''
must be created by the federation community, involving all stakeholders including the users (subjects).

Taking the parallel of an identity system, an OPAL-based information sharing system must address the following:

\begin{itemize}

\item	Verifying the correct matching between an identity (connected to a human, legal entity, device, or digital object)
and the set of data in a repository pertaining to that identity;

\item	Providing the correct result from an OPAL-based computation to the party that requires it to complete a transaction;

\item	Maintaining and protecting the private data within repositories over its lifecycle.

\item	Defining the legal framework that defines the rights and responsibilities of the
parties, allocates risk, and provides a basis for enforcement.

\end{itemize}

Similar to -- and building upon -- an identity system operating rules,
new additional operating rules need to be created for an OPAL-based information sharing system.
There are two (2) components to this.
The first is the business and technical operational rules and specifications necessary to make the OPAL-based system
functional and trustworthy.
The second is the contract-based legal rules that (in addition to
applicable laws and regulations) define the rights and legal obligations of the parties
specific to the OPAL-based system and facilitate enforcement where necessary.

As the current work is intended to focus on the concepts and principles of open algorithms and 
their application to information sharing in the identity context, these two aspects will be subject for future work.

\begin{figure*}[!t]
\centering
	% ORIGINAL \includegraphics[width=3.0in]{hardjono-classic-id-federation-v02-pdf}
	%
\includegraphics[width=0.7\textwidth, trim={0.5cm 0.5cm 0.5cm 0.5cm}, clip]{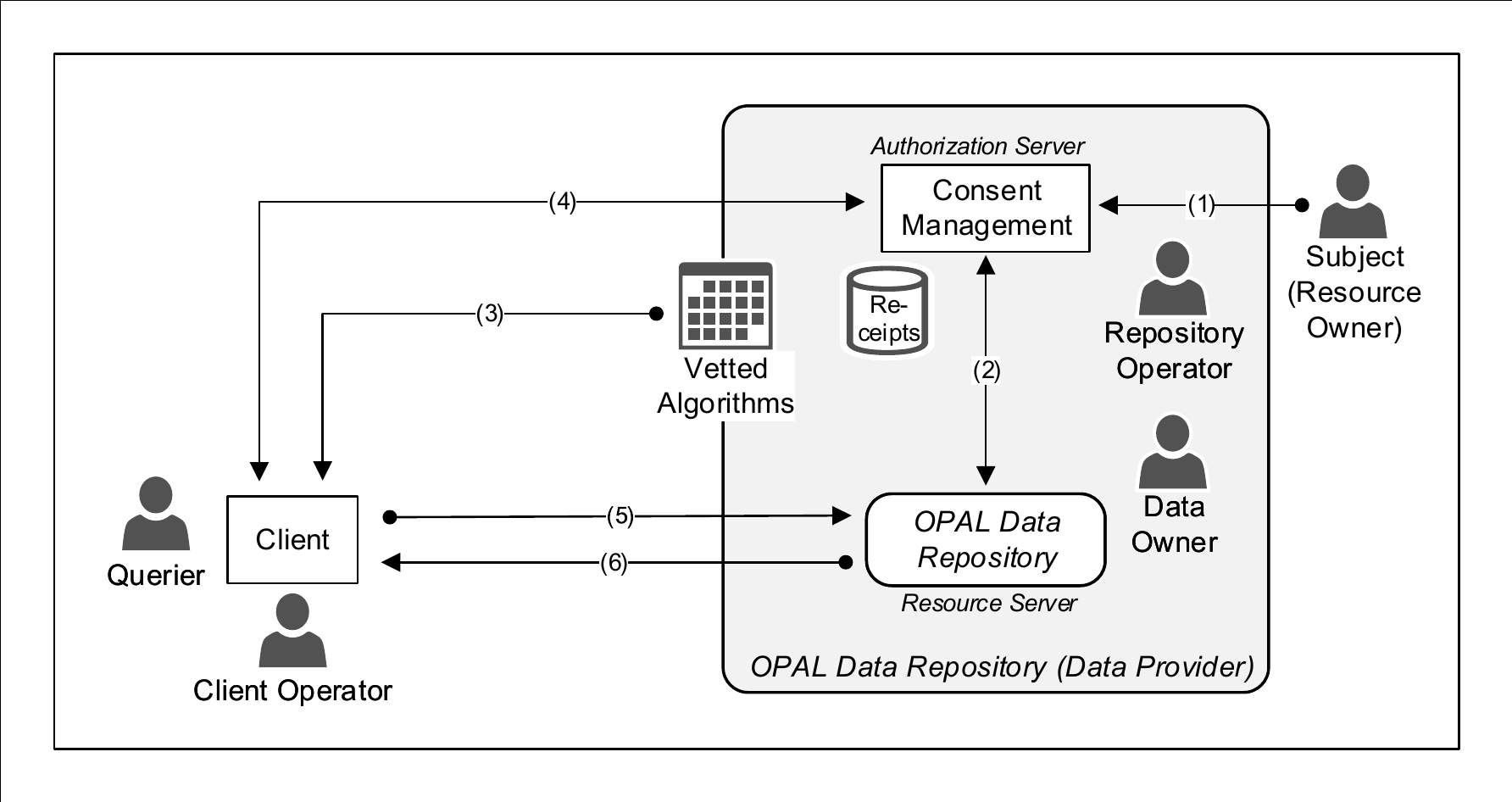}
	%
	% where an .eps filename suffix will be assumed under latex, 
	% and a .pdf suffix will be assumed for pdflatex; or what has been declared
	% via \DeclareGraphicsExtensions.
	%
	% TRIMMING:  trim={<left> <lower> <right> <upper>} and clip options:
	% FULL EXAMPLE: \includegraphics[width=0.4\textwidth, trim={0.5cm 0.5cm 0.5cm 11.3cm}, clip]{image1.pdf}
	%
\caption{Overview of Consent Management for OPAL}
\label{fig:opalconsent}
\end{figure*}

%%%% ---------- Section --------------------
\section{Consent Management in OPAL}
\label{sec:opal-Consent}

The OPAL paradigm introduces an interesting perspective on consent management,
because the subject is asked to consent to the execution
of an algorithm.
This is in contrast to the prevailing mindset today~\cite{WEF2011}
where the subject is asked permission for the data provider
to share the subject's raw data with other entities with whom
the subject did not have a transactional relationship.

Since the topic of consent is complex and outside the scope of the current paper,
we will only touch on it briefly by providing an overview
of how a consent management system could augment
an OPAL Data Provider deployment.

\subsection{Consent to Execute an Algorithm}

Asking a subject's permission to execute an algorithm on their data
and obtain ``safe answers'' is radically
different from asking the subject for permission to copy (or ``share'') their data to
an external entity.
This is true regardless
if a ``subject'' is an individual, an organization or a corporation.
The OPAL approach provides a greater degree of assurance to the subject
that the raw data-set remains where it is (one copy or few local copies)
and that the identity of the Querier is known.

In OPAL deployments 
the composition of the consent notice and receipt should include at least the following:
\begin{itemize}

\item	{\em Algorithm and algorithm identification}: 
This indicates which vetted algorithm will be executed against the data-set.
This may be a list of multiple algorithms that are designed and vetted to run against the data-set.
An implementation of OPAL may include references (e.g. URIs, hashes, etc.) that point back
to the signed template which contains the complete algorithm expressed in a given syntax.

\item	{\em Data-set identification}: Data-sets and copies of them must be
uniquely identifiable within an organization.
This could be a fixed identifier with local or global scope,
or an identifier (e.g. URI) that resolves to
an actual resource (i.e. file) containing the data-set.

\item	{\em Data provider identification}: This is the unique identity of the data provider
which holds the subject's data.
Examples include an {X.509} certificate issued by a valid Certificate Authority.
Note that more complex deployment cases may involve a {\em Repository Operator} entity
who hosts the OPAL data provider system but does not have any legal
ownership to the data.

\item	{\em Querier identification}: This is the unique identity of the querier
(e.g. {X.509} certificate).

\item	{\em Terms of use}: This is the terms of use
(or terms of service) relating to the result of the execution of the algorithm. 
A simplified easy-to-understand prose must be included for readability by the subject.
The same terms of use (legal prose) must be presented to the Querier
(e.g. included in the algorithm ``template'' description -- see Section~\ref{subsec:opal-query-response}).

\end{itemize}

\subsection{The UMA Model for Consent Management}

Given that the majority of social media platform providers today
are deploying architectures for identity management
and authorization based on the {OAuth2.0} framework~\cite{rfc6749},
it makes technological sense to extend the rudimentary {OAuth2.0} framework
for the purposes of consent management.
The {\em User Managed Access} (UMA) profile~\cite{UMACORE1.0,UMACORE2.0} of {OAuth2.0} provides such an extension.

A high-level illustration of the UMA extensions and flows
in the context of the OPAL architecture
is shown in Figure~\ref{fig:opalconsent}. 

In Figure~\ref{fig:opalconsent}, the Subject as the Resource Owner (RO)
predefines his or her consent ``rules'' at the Consent Management system in Step~1.
The Consent Management system is the UMA Authorization Server (AS).
The system matches these rules with the resources 
(e.g. files, data-sets, subsets, etc.) belonging to the Subject as the Resource Owner in  in Step~2.

After the Querier selects the Algorithm (Step~3),
the Querier must then obtain a ``consent token'' (a signed JSON data structure) from the Consent Management system  (Step~4).
The Querier binds the execution-request to the consent token by digitally signing them together,
prior to transmitting to the Data Provider (Step~5).
Finally, in Step~6 the repository returns the response to the Querier.

A key contribution of the UMA extension of {OAuth2.0} is its recognition
that in practice the Client is a web-application operated by the Client Operator (CO)
as separate legal entity from the user (Querier).
Similarly the Resource Server (RS) containing the subject's
resources may be operated
by a separate legal entity from the subject.

Note that the basic {OAuth2.0} framework~\cite{rfc6749}
does not recognize the notion of operators of services (e.g. third parties).
As such {OAuth2.0} does not distinguish between the Client web-application (to which the user connects via their browser)
from the Client Operator entity which legally owns and operates the web-application.
This is turn leads to the possibility of the operator providing
the remote web-application service without any legal obligations to the querier
or the resource owner.
More specifically, the operator of the web-application can ``listen in'' (eavesdrop) to any query/response traffic
between the querier and the data provider.
This allows the operator to obtain data and information through backdoor access via the web-application which they legally own.

In the context of OPAL in Figure~\ref{fig:opalconsent} there is an additional need to provide legal recognition
of the different entities in the data sharing ecosystem.
This includes the Subjects, the Data Owner, the Client Operator and the Repository Operator.
The Data Owner legally owns (or co-owns) the collated data, the algorithms and the information derived
from the raw data.
Individually, the Subject owns a small portion of the raw data-set in the repository.
In the case where the OPAL infrastructure is hosted,
the Repository Operator also has legal obligations (e.g. not copying or leaking raw data).

%%%% ---------- Section --------------------
\section{Future Directions}
\label{sec:opalML}

Currently there is great interest in the use of artificial intelligence (AI)
and machine learning (ML) techniques to obtain better insight
into data for various use-cases.
For the OPAL paradigm, there are several possible areas of interest:
\begin{itemize}

\item	{\em Distributed Machine Learning}: The principle of leaving raw data in their repositories
points to deployment architectures based on distributed data stores and distributed computation.
Corresponding to this architecture is the use of machine learning techniques in a distributed manner
to improve performance.

~~In an architecture with distributed instances
of OPAL data providers (data servers), one approach could be to train the algorithm
separately at each data server instance. 
Each data server instance could hold slightly different training data-sets.
The model trained at each data server instance
would then be serialized
and made available to the remote Querier.
The OPAL principles remain enforced,
where the Querier does not see the raw data but obtains the benefit
of distributed data stores performing independent training.

\item	{\em Fairness and accountability}: Fairness 
has been of concern to researchers in the area of machine learning for some time.
A key aspect of interest is in ensuring non-discrimination, 
transparency and understandability of data driven 
decision-making (e.g. see~\cite{Pentland2015,AdebayoKagal2016}).

~~For the OPAL paradigm fairness is crucial in the vetting process
for new algorithms in the context of a given data and use-case.
Transparency is a factor in obtaining consent for including data
within an given OPAL-based data federation.

\item	{\em Algorithms expressed as smart contracts}: Once an algorithm has been 
vetted to be safe to run against a given data-set, 
it can be expressed as
a smart contract for a given blockchain system or distributed ledger platform.
Here a smart contract is defined to be the combined executable code and legal-prose~\cite{NortonRoseFulbright2016}, 
digitally signed and  distributed on the P2P nodes of a blockchain system.
The legal prose would include statements on the terms of use 
for the resulting response for privacy and regulatory compliance purposes.

~~Depending on the type of blockchain system (permissioned, permissionless, semi-anonymous) 
the algorithm itself may be publicly readable. 
The querier (caller) must invoke the smart-contract 
algorithm accompanied with payment, 
which will be escrowed until the completion of 
the execution of the algorithm upon the intended data-set.

\end{itemize}

%%%% ---------- Section --------------------
\section{Conclusions}
\label{sec:furtherwork}

The identity problem of today is essentially a problem of data --
and more specifically a problem of privacy-preserving data sharing.

This paper has described the concepts and principles
underlying the open algorithms (OPAL) paradigm.
The OPAL paradigm offers a possible way forward
for industry and government
to begin addressing the core issues around privacy-preserving data sharing.
Some of these challenges include siloed data,
the limited type/domain of data, and the prohibitive situation
of cross-organization sharing of raw data.

Instead of sharing fixed-attributes regarding a user or subject,
the OPAL paradigm offers a way for Identity Providers, Relying Parties and Data Providers
to share vetted algorithms.
This in turn provides better insight into the user's behavior,
with their consent.
It also allows for the development of a trust network ecosystem
consisting of these entities,
providing new revenue sources,
governed by relevant legal agreements and contracts
that form the basis for a information sharing legal trust framework.

Finally, a new set of legal rules and system-specific rules
must be devised that must clearly articulate the required combination of technical
standards and systems, business processes and procedures, and legal rules that, taken
together, establish a trustworthy system for information sharing in a federation based on the OPAL model.

\section*{Acknowledgment}

The authors thank the following for inputs and insights (alphabetical):
Abdulrahman Alotaibi,
Stephen Buckley,
Raju Chithambaram,
Keeley Erhardt, 
Indu Kodukula,
Emmanuel Letouzé,
Eve Maler, 
Carlos Mazariegos,
Yves-Alexandre de Montjoye,
Ken Ong,
Kumar Ramanathan,
Justin Richer, 
David Shrier, 
Charles Walton.
%%%We also thank the reviewers for valuable suggestions on improvements for the paper.



%\bibliographystyle{IEEEtran}

%===== \begin{thebibliography}{1}

%===== \bibitem{IEEEhowto:kopka}
%===== H.~Kopka and P.~W. Daly, \emph{A Guide to \LaTeX}, 3rd~ed.\hskip 1em plus
%=====   0.5em minus 0.4em\relax Harlow, England: Addison-Wesley, 1999.

%===== \end{thebibliography}

%%%\bibliographystyle{IEEEtran}
%%%\bibliography{IEEEabrv,mergedbib,hardjonobib,rfcbib}


\end{document}